\documentclass[aps,twocolumn,showpacs,amsmath]{revtex4}
\usepackage{graphicx}
\usepackage{amsmath}
\usepackage{lipsum}

\begin{document}

\title{Jackiw-Rebbi-type bound state carrying fractional fermion parity}

\author{Ye Xiong} 
\affiliation{Department of Physics and Institute of Theoretical Physics
  , Nanjing Normal University, Nanjing 210023,
P. R. China}
\email{xiongye@njnu.edu.cn}
\author{Peiqing Tong}
\affiliation{Department of Physics and Institute of Theoretical Physics
  , Nanjing Normal University, Nanjing 210046,
P. R. China}
\affiliation{Jiangsu Key Laboratory for Numerical Simulation of Large
  Scale Complex Systems, Nanjing Normal University, Nanjing 210023,
P. R. China}
\affiliation{Kavli Institute for Theoretical Physics China, CAS,
Beijing 100190, China} 
\email{pqtong@njnu.edu.cn}

\begin{abstract} 
  We find the coexistence of two kinds of non-abelian anyons, Majorana
  fermion at the geometric ends and Jackiw-Rebbi-type bound state (JRBS)
  at a domain-wall, in a new topological superconducting phase in
  one-dimensional (1D) systems.  Each localized JRBS carries a new
  fractional quantity, the half of the parity of fermion number. This
  induces a topological protected crossing at the zero energy for its
  eigen-energy.  For a chain embedded with a JRBS, one is possible to
  switch between the occupied and the empty states of Majorana zero energy
  state (MZES) by varying the strength of external magnetic field across
  that crossing point. This enable a way to encode a quantum qubit into
  one MZES without breaking parity conservation. We propose that such
  JRBS and Majorana fermion can appear in two 1D models, one can be
  accomplished in an artificial lattice with staggered hopping,
  staggered spin-orbital interaction and staggered superconducting
  pairing in cold fermion atoms, the other is a 1D
  semiconductor chain sandwiched between s-wave superconductor and
  antiferromagnet.
\end{abstract}

\maketitle

\section{Introduction} 

It has been proposed that Majorana Fermions (MF)
can exist in a topological superconducting phase (TSP) at the core of 
magnetic vortex penetrating a 2-dimensional
(2D) $p_x \pm ip_y$ superconductor\cite{Ivanov2001, RevModPhys.83.1057,
RevModPhys.80.1083, PhysRevLett.112.086401, PhysRevLett.100.096407,
PhysRevB.82.184516, PhysRevB.84.054502, oddperiod,
PhysRevLett.112.037001} or at the ends of a 1-dimensional
(1D) $p$-wave superconductor \cite{Kitaev, PhysRevLett.104.040502,
TIbook, PhysRevLett.109.236801, PhysRevLett.105.227003,
PhysRevLett.100.096407, PhysRevB.90.014505}. These MFs, being their own anti-particles,
obey non-abelian braiding statistics so that the quantum computing based
on them is fault-tolerant \cite{RevModPhys.80.1083, RevModPhys.82.3045}.
In practice, a quantum qubit is encoded into two Majorana zero energy
states (MZES) but not into one because the parity conservation
prevents the switching between the occupied and the empty states of
a MZES. This restriction definitely increases the complexity of the
topological computation in experiment. 

Besides MF, there is another kind of topological impurity in 1D
system, the Jackiw-Rebbi-type bound
state (JRBS) on a domain-wall\cite{PhysRevD.13.3398, TIbook}. Su
and et al. had studied the tight-binding model of polyacetylene, now
known as the Su-Schrieffer-Heeger (SSH) model, and found that the
soliton state on the domain wall was JRBS\cite{PhysRevLett.42.1698,
PhysRevB.22.2099}. One of the exotic behaviors of the JRBS is that it
carries fractional charge $e/2$ \cite{PhysRevLett.42.1698,
PhysRevB.22.2099, PhysRevB.25.6447}. This is the first quasiparticle,
in the single-particle picture, that possesses only a fraction of the
elementary charge $e$. But in polyacetylene, this fractional charged
soliton can not be observed because the spin degeneracy makes the two
fractional charges in the two spin subspaces compensate to $e$ or $0$.
There are many proposals to lift this spin degeneracy for observing the
fractional charge carried by JRBS \cite{PhysRevLett.108.136803,
PhysRevLett.112.196803, PhysRevLett.107.166804,
PhysRevLett.109.236801}.

Furthermore, JRBS is also a kind of anyon obeying non-abelian
braiding statistics \cite{PhysRevLett.110.126402}. One may image a
situation with both MF and JRBS and braiding them together. It has been found
that the bound states at the geometric ends can change from JRBS to MF
\cite{PhysRevLett.109.236801}. But a 1D system that can intrinsically    
host both MF and JRBS has not been found.

\begin{figure}[ht]
  \includegraphics[width=0.45\textwidth]{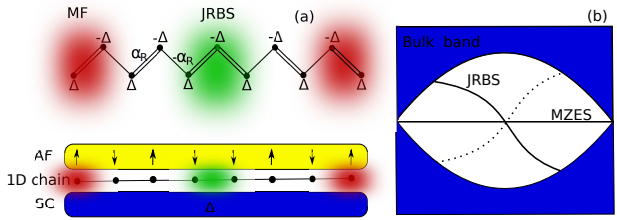}
  \caption{(Color online) (a) two 1D models we studied. In the upper model, the double
    bonds and single bond are used to illustrate the alternative
    stronger and weaker hoppings between the nearest neighboring sites.
    The spin-orbital interaction $\alpha_R$, as well as the on-site
    Cooper pairing $\Delta$ are also staggered along the chain. The
    domain-wall, simulated by two adjacent stronger bonds, can host one
    JRBS (green cloud), while MF appears at the geometrical ends(red
    cloud). In the bottom model, the s-wave superconducting pairing and
    antiferromagnetic (AF) order are introduced to a uniform
    semiconductor chain by the proximity effect.  The domain-wall is
    simulated by an AF domain-wall. (b) A schematic illustration of the
    energy spectrum of bulk states, the eigen-energy of JRBS and MZES in
    the TSP. The blue regions represent the bulk band. The JRBS must
    continuously connect the particle and hole bands and inevitably go
    through zero energy at a point. The dotted line shows the
  eigen-energy of its antiparticle obeying the particle-hole symmetry.} 
  \label{fig1} 
\end{figure}

In this paper, we raise two 1D models that can host MF and JRBS
simultaneously. We are able to switch between the occupied and empty
states of a MZES by varying external magnetic field.  This
manipulation depends on a crossing at the zero energy for the
eigen-energy of JRBS, which has been schematically showed in Fig.
\ref{fig1}(b). This crossing is topologically protected so that the
manipulation is robust against local disorder.

At the first glance, it seems surprising that a localized JRBS can
affect the global properties encoded in MZES. The key clue is that in
the presence of superconducting coupling, JRBS has abandoned one of its
famous properties: each JRBS carries fractional charge $e/2$. This is
due to the broken of fermion number conservation. But the parity
conservation of fermion number is still present which makes JRBS carry
fractional parity(FP), a fractional quantity used to hide behind
fractional charge in the nonsuperconducting models. It is in this way
that the localized JRBS links with the global property, parity of total
fermion number. 

First of all, we want to illustrate how FP occurs in the 1D systems.
Suppose there are two infinite chains, A and B.  A is uniform and B has
a pair of long separating JRBSs and is uniform elsewhere.  The
parameters on A and B are the same. The two JRBSs on B are far from each
other so that each one can be considered individually. In the absence of
superconducting pairing, the total numbers of fermions are well defined,
denoted as $N_A$ and $N_B$ in the chains $A$ and $B$ respectively. A
standard Thouless pump tells us that the two JRBSs in B induce a
relation, $|N_A -N_B|=1$. The fractional charge $e/2$ carried by each
JRBS is produced in this argument because each JRBS must take the
responsibility of the half of one elemental charge caused by the fermion
number difference.  When the superconducting pairing is nonzero, the
conserved quantities on the chains, A and B, regress from fermion number
to fermion parity, $P_{A(B)} = N_{A(B)} \, \text{mod} \: 2$.  We will
show that the well defined (conserved) quantities on A and B are
different by $|P_A -P_B|=1$. So each JRBS in a superconducting model
takes the responsibility of the half of one parity difference. This is
the source from where the concept, FP, comes.

We will show that these features could be realized in two 1D systems
showed in Fig. \ref{fig1} (a). The first model can be realized with cold
atoms in an artificial 1D lattice with staggered nearest neighboring hopping, staggered
spin orbital interaction and staggered superconducting pairing. The
latter one is more easier to be carried out by sandwiching a
semiconductor chain between an antiferromagnet(AF) and an ordinary
s-wave superconductor. In our numerical calculation, the domain-wall is
simulated by two adjacent stronger(weaker) bonds in the first model and
by an AF domain-wall in the latter one. But our conclusions, in general,
do no depend on the actual size and shape of the domain-walls. 

In section 2, we will concentrate on the first model. Its phase diagram,
the FP JRBS, the coexistence of JRBS and  MF, the unavoidable zero
energy crossing for JRBS and how to encode a qubit into one MZES with
the help of a JRBS are discussed in this section. In section 3, we study
paralleled on the second model. Section 4 is the conclusions. 

\section{The first model}

\subsection{The Hamiltonian of the first model} 

We start from a theoretical 1D tight-binding Hamiltonian,
\begin{eqnarray}
  H & = &\sum_{i\beta} \mu c^\dagger_{i\beta} c_{i\beta}+
  \sum_{i\beta\gamma} [1-(-1)^i\delta]  (c^\dagger_{i+1\beta} \sigma^z_{\beta\gamma}
  c_{i\gamma} +\text{h.c.}) \nonumber \\
  & & +\sum_{i\beta\gamma} Bc^\dagger_{i\beta} \sigma^z_{\beta\gamma}
  c_{i\gamma} 
  +\alpha_R \sum_i (c^\dagger_{i\uparrow}c_{i+1\downarrow} -
  c^\dagger_{i\downarrow}c_{i+1\uparrow}+ \text{h.c.}) \nonumber \\
  & & +\sum_{i} \Delta (c^\dagger_{i\uparrow}c^\dagger_{i\downarrow}
  +\text{h.c.}).
  \label{Ham1}
\end{eqnarray}
Here, $c_{i\beta}$ and $c^\dagger_{i\beta}$ are the annihilation and
creation operators for spinful fermion with spin $\beta$ on site $i$ and
$\sigma$'s are Pauli matrices. The strength of hopping between the
nearest neighboring sites stagger between $1+\delta$ and $1-\delta$,
where the energy unit is set as the uniform part of hopping strength.
Each unit cell contains the sites from the two sublattices, denoted by
$A$ and $B$, respectively. $\sigma^z$ appears in the hopping term
because we have applied a transformation, $c_{(2n+1)\downarrow} \to -
c_{(2n+1)\downarrow}$, on the odd sites of the lattice for the upper
model showed in Fig. \ref{fig1}(a). The parameters $\mu$, $\delta$,
$\alpha_R$ and $\Delta$ are for the strength of the chemical potential,
the staggered part of hopping, the staggered spin-orbital interaction
and the staggered superconducting pairing, respectively.

Such Hamiltonian, Eq. \ref{Ham1}, may be realized with cold fermions
trapped in a 1D laser induced lattice.  The staggered hoppings like that
in the SSH model has been realized in the experiment\cite{catomssh,
atom}. In a recent proposal \cite{PhysRevLett.112.086401}, the staggered
effective spin-orbital interaction can also be produced with the aid of
modern technologies. It was also known that 1D Fermi gas with spin
orbital coupling was dominated by Fulde-Ferrell (FF) superfluid phase at
the low temperature \cite{ PhysRevLett.108.225302,
PhysRevLett.112.136402, PhysRev.135.A550, PhysRevLett.111.235302}. This
FF phase, if properly choosing the lattice constant of the 1D lattice,
can be simulated with a staggered pairing coefficient.  So the
tight-binding Hamiltonian of the system reads 
\begin{eqnarray}
  H & = & \sum_{i\beta} \mu c^\dagger_{i\beta} c_{i\beta}+
  \sum_{i\beta} [ 1-(-1)^i\delta ]
  (c^\dagger_{i+1\beta}c_{i\beta}+\text{h.c.}) \\ \newline \nonumber
  & & +\sum_{i\beta\gamma} Bc^\dagger_{i\beta} \sigma^z_{\beta\gamma} 
  c_{i\gamma} + \sum_{i} (-1)^i\Delta
  (c^\dagger_{i\uparrow}c^\dagger_{i\downarrow}
    +\text{h.c.}) \\ \newline \nonumber
   & & - \sum_i(-1)^i\alpha_R (c^\dagger_{i\uparrow}c_{i+1\downarrow} 
  + c^\dagger_{i\downarrow}c_{i+1\uparrow}+\text{h.c.}),
  \label{Ham1_r}
\end{eqnarray}
where the chemical potential, the staggered hoppings, the magnetic field
induced Zeeman term,  the FF superfluid pairing and the staggered
spin-orbital interaction are written, subsequently. Through a
transformation on the odd lattice, $c_{2n+1\uparrow} \to
c_{2n+1\uparrow}$ and $c_{2n+1\downarrow} \to -c_{2n+1\downarrow}$, the
staggered spin-orbital and superconducting interactions are smeared out
in the new representation and the Hamiltonian changes to the effective
one in Eq.  \ref{Ham1}. 

\subsection{The phase diagram}

We can study the model with the periodic boundary condition so that the wave
vector $k$ is a good quantum number. The Hamiltonian in the Nambu, the spin
and the sublattice representation $(\psi_{kA\uparrow}, \psi_{kB\uparrow},
\psi_{kA\downarrow}, \psi_{kB\downarrow}, \psi^\dagger_{-kA\uparrow},
\psi^\dagger_{-kB\uparrow}, \psi^\dagger_{-kA\downarrow},
\psi^\dagger_{-kB\downarrow})^T$ reads
\begin{equation}
  H(k)=\begin{pmatrix} H_0(k) & V(k) \\ V^\dagger(k) & -H_0(k) \end{pmatrix},
  \label{Ham0k}
\end{equation}
where 
\begin{widetext}
\[
  H_0(k)=\begin{pmatrix} B+\mu & (1+\delta)+(1-\delta)e^{-ik} & 0 & \alpha_R
    (1-e^{-ik}) \\
    (1+\delta)+(1-\delta)e^{ik} & B+\mu & -\alpha_R(1-e^{ik}) & 0 \\
    0 & -\alpha_R(1-e^{-ik}) & -B+\mu & -[(1+\delta)+(1-\delta)e^{-ik}] \\
    \alpha_R(1-e^{ik}) & 0 & -[(1+\delta)+(1-\delta)e^{ik}] & -B+\mu
  \end{pmatrix}
\]
\end{widetext}
and 
\[
  V(k)=\begin{pmatrix} 0 & 0 & \Delta & 0 \\
    0 & 0 & 0 & \Delta \\
    -\Delta & 0 & 0 & 0 \\
    0 & -\Delta & 0 & 0 \end{pmatrix}.
\]

Through a unitary transformation 
\[U=\frac{1}{\sqrt{2}}\begin{pmatrix} I & I \\ I & -I \end{pmatrix}, \]
the Hamiltonian is transformed to 
\[ H(k) \to U H(k) U^{-1} = \begin{pmatrix} 0 & A(k) \\
  A^\dagger(k) & 0 \end{pmatrix}, \] where $I$ is a $4\times 4$ unit matrix
and $A(k)= H_0(k) +V(k)$. 

For a gapped ring, the band gap can only close at $k=0$ or $k=\pi$ in the
Brillouin zone as varying parameters. At these phase boundaries, the
nonzero bulk wavefunction at $E=0$ implies $\text{det}(A)=0$. So we
have the two phase boundary conditions, $B^2= \Delta^2+\mu^2+4\pm
4\sqrt{\Delta^2+\mu^2}$ from $k=0$ and
$B^2=\Delta^2+4\delta^2+\mu^2-4\alpha^2\pm
4\sqrt{\Delta^2\delta^2+\delta^2\mu^2-\Delta^2\alpha^2}$ from $k=\pi$.

\begin{figure}[ht]
  \includegraphics[width=0.45\textwidth]{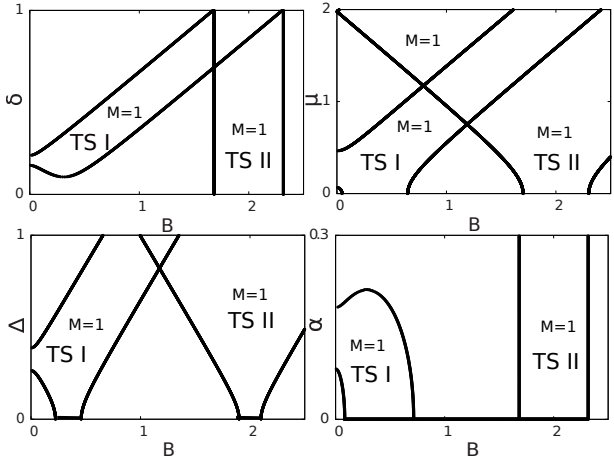}
  \caption{The phase diagram in $B-\delta$ (a) , $B-\mu$ (b), $B-\Delta$
    (c) and $B-\alpha_R$ (d) planes. The other parameters
    are $\mu=0.1$, $\alpha_R=0.1$ and $\Delta=0.3$ in (a),
    $\alpha_R=0.1$, $\Delta=0.3$ and $\delta=0.2$ in (b),
    $\alpha_R=0.1$, $\delta=0.2$ and $\mu=0.1$ in (c) and
    $\delta=0.2$, $\mu=0.1$ and $\Delta=0.3$ in (d).
    The regions in the TSP with a topological invariant $M=1$ have been
    indicated explicitly. The rest regions are for topological trivial
  phase with $M=0$. The coexistence of MF and JRBS happens only in the
  TPS indicated by ``TS I''.}
  \label{fig2_1}
\end{figure}

In Fig. \ref{fig2_1}, we sketch the phase diagram in $B-\mu$,
$B-\alpha$, $B-\Delta$ and $B-\delta$ planes, respectively by
numerically diagonalizing $H(k)$. The phase boundaries are consistent
with the above two conditions except on the $B$ axis when $\alpha=0$ or
$\Delta=0$.  This deviation is because in these particular conditions,
the model is gapless, which violates our assumption that the gap closes
at $k=0$ or $k=\pi$.

A topological invariant can be defined by $M= \frac{\Phi_{ZB}}{\pi} \,
\text{mod} \, 2$, where $\Phi_{ZB}$ is the Zak-Berry phase integrated
over the whole Brillouin zone $\Phi_{ZB} = \int_{-\pi}^{\pi} -i \langle
\psi | \partial_k | \psi \rangle dk$ and $|\psi \rangle $ is the
eigenstate with the negative energy at $k$.  The above topological
invariant specified by the Zak-Berry phase is equivalent to the Pfaffian
invariant first introduced by Kitaev in studying 1D topological
superconductor \cite{PhysRevB.88.075419}.  In Fig. \ref{fig2_1}, we
indicate the regions in the topological superconducting phase (TSP) with
$M=1$. The rest regions are for the topological trivial phase with
$M=0$.  We will show that, the phase diagram contains two kind of TSPs,
denoted by ``TS I'' and ``TS II'' in the figures, respectively.
MFs and JRBS can only coexist in ``TS I'', a region existes only when
$|\delta|>|\alpha_R|$, but not in ``TS II'' stemming from $B=2$.

\begin{figure}[ht]
  \includegraphics[width=0.45\textwidth]{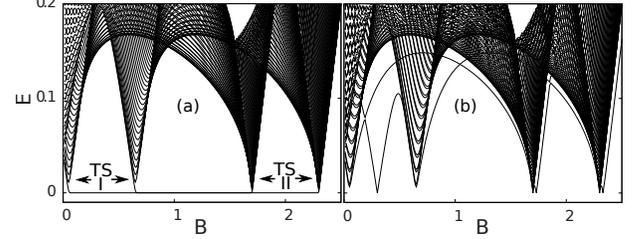}
  \caption{The energy spectrum for $N=400$ chain with open boundary
    condition (a) and for $N=401$ ring with periodic boundary condition
    (b). Only the positive eigenenergies are shown.
    Parameters are $\delta=0.2$, $\Delta=0.3$, $\alpha_R=0.1$  and
    $\mu=0$. In (a), a single MZES appears in two regions:
    ``TS I'' $0.05<|B|<0.65$ and ``TS II'' $1.7<|B|<2.3$. There are two
    MZESs in the region $0.65<|B|<1.7$. Panel (b)
    shows that a domain-wall can bring {\it one} bound state deep in
  band gap in ``TS I''.}
  \label{fig2_2}
\end{figure}

These topological nontrivial phases can be confirmed by the existence of
boundary states at the geometrical ends.  In Fig. \ref{fig2_2}(a), we
plot energy spectrum for the Hamiltonian with open boundary condition.
The length of the chain is $N=400$ and the chemical potential is
$\mu=0$. If not mentioned, in this paper, the spectrum show only the
eigenenergies with positive energies. Their counterparts with negative
energies are not explicitly shown. 

Fig. \ref{fig2_2} (a) shows that there is one Majorana zero energy state
(MZES) in the band gap in two regions: ``TS I'' in $0.05 < |B| < 0.65$
and ``TS II'' in $1.7<|B| < 2.3$. There is also another exotic region in
$0.65<|B|<1.7$, where two MZESs appear.  The double-degenerate Kramers
MF bound states have been discussed in a two-chains model with
particle-hole and time-reversal symmetry in Ref.
\cite{PhysRevLett.112.126402, PhysRevB.89.220504}.  The two zero-energy
bound states in our model are similar to this MZES pair but the
time-reversal symmetry has been replaced by the sublattice symmetry when
$\mu=0$.

In Fig.  \ref{fig2_2}(b), we plot the energy spectrum for the model with
periodic boundary condition and the length of the ring is changed to
$N=401$.  As the length of the unit cell is $2$, the ring contains
insuppressible half unit cell. So this ring naturally engages a
domain-wall and the energy spectrum exhibit the bound state at the wall.
In Fig. \ref{fig2_2}(b) MZES disappears as there is no geometric end.
Outside ``TS I'', the energies of bound states are adjacent to the bulk
band, implying that the domain-wall can only be considered as a normal
impurity in that case. In ``TS I'', however, a bound state deep-in-gap
can evolve continuously across the zero energy. This implies that the
domain-wall in ``TS I'' should be considered as a topological impurity
that triggers {\it one} JRBS. 

\subsection{Another way to understand TSP when $\mu=0$}
When $\mu=0$, through a unitary transformation 
\[
  U=\frac{1}{\sqrt{2}}\begin{pmatrix} 
    1 & 0 & 0 & 0 & 0 & 0 & -1 & 0 \\
    0 & 1 & 0 & 0 & 0 & 0 & 0 & -1 \\
    0 & 0 & 1 & 0 & -1 & 0 & 0 & 0 \\
    0 & 0 & 0 & 1 & 0 & -1 & 0 & 0 \\
    0 & 0 & 1 & 0& 1 & 0 & 0 & 0 \\
    0 & 0 & 0 & 1& 0 & 1 & 0 & 0 \\
    1& 0 & 0 & 0 & 0 & 0 & 1 & 0 \\
    0 & 1& 0 & 0 & 0 & 0 & 0 & 1 \\ \end{pmatrix},
\]
the Hamiltonian can be decoupled into two partitioning parts
\begin{equation}
  H\to U H U^\dagger =\begin{pmatrix} H_+ & 0 \\
    0 & H_- \end{pmatrix},
  \label{Hpa}
\end{equation}
where 
\begin{widetext}
\[
  H_- = -\begin{pmatrix} B-\Delta & (1+\delta)+(1-\delta)e^{-ik} & 0 & \alpha_R
    (1-e^{-ik}) \\
    (1+\delta)+(1-\delta)e^{ik} & B-\Delta & -\alpha_R(1-e^{ik}) & 0 \\
    0 & -\alpha_R(1-e^{-ik}) & -B+\Delta & -[(1+\delta)+(1-\delta)e^{-ik}] \\
    \alpha_R(1-e^{ik}) & 0 & -[(1+\delta)+(1-\delta)e^{ik}] & -B+\Delta
  \end{pmatrix}
\]
and 
\[
  H_+ = \begin{pmatrix} B+\Delta & (1+\delta)+(1-\delta)e^{-ik} & 0 & \alpha_R
    (1-e^{-ik}) \\
    (1+\delta)+(1-\delta)e^{ik} & B+\Delta & -\alpha_R(1-e^{ik}) & 0 \\
    0 & -\alpha_R(1-e^{-ik}) & -B-\Delta & -[(1+\delta)+(1-\delta)e^{-ik}] \\
    \alpha_R(1-e^{ik}) & 0 & -[(1+\delta)+(1-\delta)e^{ik}] & -B-\Delta
  \end{pmatrix}.
\]
\end{widetext}

\begin{figure}[ht]
  \includegraphics[width=0.45\textwidth]{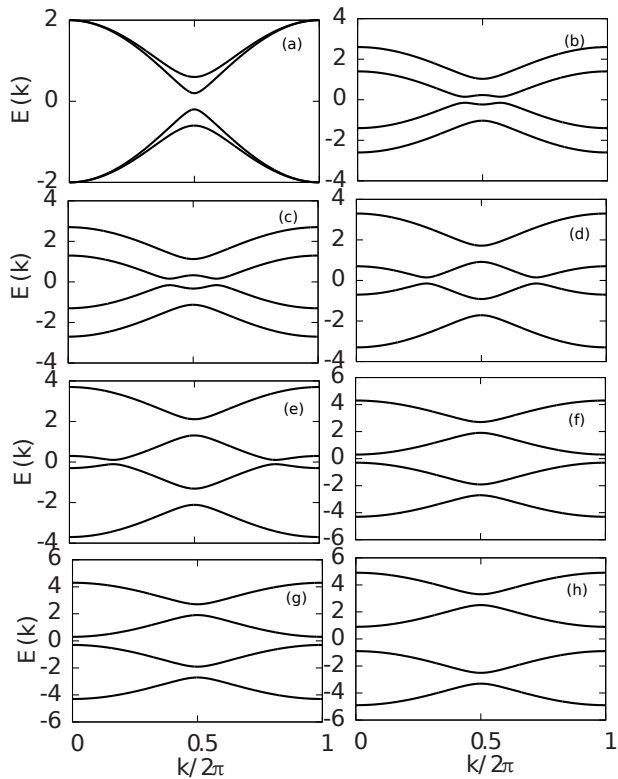}
  \caption{The band dispersion with $k$ for the two partial
    Hamiltonians, $H_-$ (a) (c) (e) (g) and $H_+$ (b) (d) (f) (h). The
    parameters are $\alpha_R=0.1$, $\delta=0.2$, $\Delta=0.3$ and
  $B=0.3$ (a) (b), $B=1$ (c) (d) $B=2$ (e) (f) and $B=2.6$ (g) (h).}
  \label{fig2_3}
\end{figure}

We show the dispersion of the eigen-energies for the two partitioning
parts, $H_-$ and $H_+$, in different phases in Fig. \ref{fig2_3},
respectively.  (a), (c), (e) and (g) are for $H_-$ and (b), (d), (f),
(h) are for $H_+$.  The four rows of panels show the dispersion with
$B=0.3$ (in ``TS I''), $B=1$, $B=2$ (in ``TS II'') and $B=2.6$,
respectively. The band inversion happens only in one partitioning part
of the Hamiltonian in ``TS I'' and ``TS II''. This is consistent with our
conclusions that ``TS I'' and ``TS II'' are in the TSP with only one MZES.  The
region in between I and II can host totally two MZESs, one in $H_+$ and
the other in $H_-$.

\subsection{Fractional parity JRBS}

Next, we will use a topological argument to prove that each JRBS carries
FP. From this, we can conclude that the zero energy crossing for JRBS is
unavoidable. After that, the application of this property on the
controllable switching of the occupation states of a MZES is presented.

We use the evolution of Wannier functions (WF) during the Thouless pump
to complete a topological proof of the assertion raised in the
introduction.

We extend the Thouless pump (charge pump), first introduced to the SSH
model \cite{TIbook}, to the present spinful model.  It is introduced by
modifying the Hamiltonian with an extra parameter $\phi$,
$H(\phi)=H_0(\phi)+H_\text{st}(\phi)$, where $H_\text{st}(\phi) = \sum_i
h_\text{st}\sin(\phi) (-1)^i (c^\dagger_{i\uparrow}c_{i\uparrow}-
c^\dagger_{i\downarrow} c_{i\downarrow})$ and $H_0(\phi)$ is a modified
Hamiltonian by replacing $\delta$ with $\delta \cos(\phi)$ in Eq.
\ref{Ham1}. The absolute value of $h_\text{st}$ is moderate so that the
band gap at the Fermi energy is not closed during the pump.

The most localized WFs \cite{ThoulessWannier, PhysRevB.74.235111,
PhysRevLett.107.126803} for the occupied bands are obtained from the
eigenvectors of the tilde position operator $\tilde R(\phi) = \hat
P(\phi) \hat R \hat P(\phi) $, where $\hat R$ is the position operator
extended to the Nambu representation and $\hat P(\phi) = \sum_{\alpha
\in \text{occupied states}} |\alpha(\phi)\rangle \langle
\alpha(\phi)|$ is the project operator on the occupied states ($E<0$)
for the Hamiltonian $H(\phi)$.  Here the position operator is $\hat R
=\text{diag}(1,2,\cdots,N)\tau_0$, where $\tau_0$ is the $2\times2$
unit matrix in the particle-hole subspace and
$\text{diag}(1,2,\cdots,N)$ is a diagonal matrix with the diagonal
elements running through lattice sites from $1$ to $N$.  The
eigenvalues of $\tilde R(\phi)$, denoted as $R$s, are the central positions
of the WFs. It should be noticed that in the Nambu representation,
each unit cell contributes $4$ WFs while in a half filled spinless SSH
model, it contributes only $1$ WF.

\begin{figure}[ht]
  \includegraphics[width=0.45\textwidth]{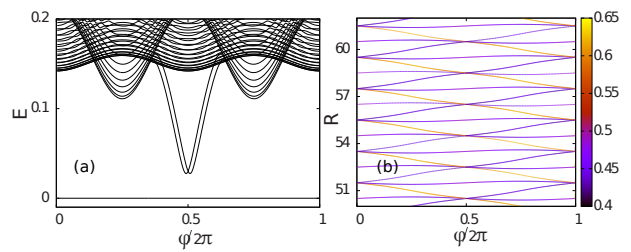}
  \caption{The energy spectrum for $H(\phi)$ with open boundary
    condition (a) and the associated center positions of WFs (b).The
    length of the chain is $N=400$.  Parameters are $\delta=0.2$,
    $\Delta=0.3$, $\mu=0.1$ $\alpha_R=0.1$ and $h_\text{st}=0.3$. The
    color palette in (b) is indicating the weights of WFs projected onto
  the particle subspace in Nambu representation.}
  \label{fig2_4}
\end{figure}

In Fig. \ref{fig2_4}, we plot the energy spectrum (a) and the center
positions of
WFs (b) during the Thouless pump with the parameters in ``TS I''. The
energy spectrum shows that with the moderate value of $h_\text{st}=0.3$,
the band gap keeps open during the pump. This fact ensures that the WFs
are localized and their center positions showed in (b) are reliable
\cite{PhysRevB.26.4269}.  According
to the evolution of the center positions of WFs showed in Fig.  \ref{fig2_4}(b), these
WFs can be classified into two groups, one corresponding to the WFs that
do not change their position after a circle of pump and the other
corresponding to the WFs that change their positions by one unit cell.
The WFs in the latter group can be further divided into two kinds,
one(in blue) is those moving in the positive direction with $\phi$ and
the other (in red) includes those moving inversely. 

\begin{figure}[ht]
  \includegraphics[width=0.45\textwidth]{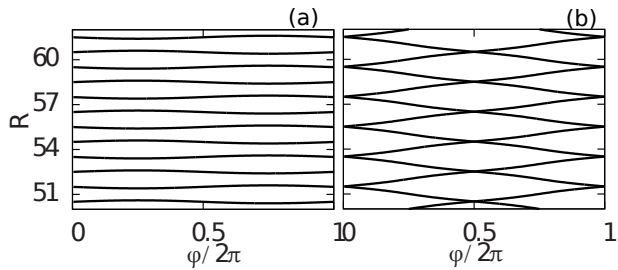}
  \caption{The evolution of the center positions of WFs during the
    Thouless pump for the two partitioning Hamiltonians, $H_+$ (a) and $H_-$
    (b). The parameters are same as those in Fig. \ref{fig2_4} except
  $\mu=0$.}
  \label{fig2_5}
\end{figure}

In the above subsection, we show that the Hamiltonian can be decoupled into
two parts, $H_\pm$, when $\mu=0$. Increasing $\mu$ from $0$ prohibits
this decoupling but the topological properties of the band keep
invariant until the band gap closes. After compared Fig. \ref{fig2_4}
(b) with the evolution of the center positions of the WFs for the
partitioning Hamiltonian $H_\pm$, shown in Fig. \ref{fig2_5}, 
we can conclude that the above two groups of WFs inherit the
evolution with $\phi$ from those of the partial Hamiltonians $H_\pm$,
respectively. The WFs inherited from those of $H_+$ experience a trival
evolution (WFs come back to their initial positions) after a circle of
pump while the other set that undergo a nontrivial evolution (WFs switch
one unit cell) come from $H_-$. If transforming $H_{\pm}$ back
to the lattice representation through an inverse Fourier transformation,
one can find MF at the ends in $H_+$ and JRBS at domain-wall in $H_-$. 

We have also studied the evolution of the center positions of WFs with
the parameters in ``TS II''. But the WFs do not show any nontrivial
evolution in that case.  

Fig. \ref{fig2_4}(b) can help us recognize that each JRBS carries FP.
The topological proof includes $4$ steps and we would like to highlight
the goal of each step at the first. In the 1st step, besides the chains
A and B raised in the introduction, an auxiliary chain C is employed. C
is not uniform but with the pump parameter $\phi$ varying slowly along
it from $0$ to $2\pi$. The other parameters are the same as those in the
uniform chain A. From the evolution of WFs showed in Fig.  \ref{fig2_4}
(b), on account of the total numbers of WFs, we can conclude that chains
C and A are different by one pair of WFs. In the 2nd step, we prove that
chains B and C have the same numbers of WFs. So with the bridge: chain
C, we find that chains A and B are different by the pair of WFs in the
number of total WFs. In the 3rd step, at a particular set of parameters,
$\mu=0$, $B=0.3$ and $\alpha_R=0$, the pair of WFs implies that the
total number of quasi-particles in A and B are different by one in the
representation of $H_{\pm}$. In the 4th step, after coming back to the
original Nambu representation, the above one quasi-particle difference
is equivalent to the parity difference between chains A and B. When the
parameters leave away from these particular ones, the above conclusion
is not modified as long as they are still in ``TS I''.

\begin{figure}[ht]
  \includegraphics[width=0.45\textwidth]{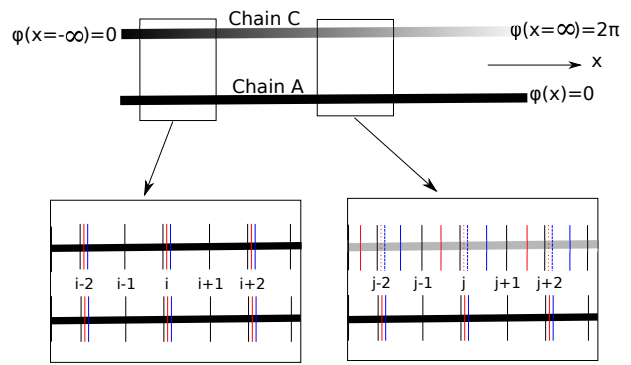}
  \caption{Chain C is describing by the extended Thouless pump Hamiltonian
  $H(\phi(x))$ with $\phi(x)$ slowly varying along the chain. The short
vertical lines in the two zoomed pictures represent the localized WFs.
There are 4 WFs in each unit cell and $i$,$j$ are used to denote the  
positions along the chain.}
  \label{fig2_6}
\end{figure}

{\it The first step.---} Let us compare the center positions of WFs in
the two infinite chains A and C, which are described by the Thouless
pump Hamiltonian $H(\phi)$. Chain A is a uniform chain with $\phi=0$
(the Hamiltonian regresses to Eq. \ref{Ham1}) and B is a chain on which
$\phi$ is slowly varying $2\pi$ along it, which have been schematically
showed in Fig. \ref{fig2_6}.  Without loss of generality, we let
$\phi(x=-\infty)=0$ and $\phi(x=\infty)=2\pi$ in C.  Because $\phi(x)$
is varying very slowly, in each macroscopically small but
microscopically large segments, it can be considered as a constant. This
ensures us to find the positions of the WFs in each segment in the
latter chain. For the segments at $x=-\infty$, the positions of WFs in
two chains are identical.  But as $x$ is increasing, compared with those
in A, a set of WFs (in blue) in C begins to misalign slightly in the
positive $x$ direction while another set of WFs (in red) has misaligned
simultaneously in the negative direction, as showed in the figure.  As
we sweeping our focus through the chains from $x=-\infty$ to
$x=+\infty$, the above misalignments increase and finally reach $\pm2$,
the length of a unit cell. This can be considered as that tunning on
$\phi(x)$ in chain C will push a WF (in blue) outside and pull a WF (in
red) inside at $x=+\infty$.  So we can conclude that compared with the
uniform chain, C donates one WF (in blue) and accepts another WF (in
red). 

{\it The second step.---} Now we relax the restriction that $\phi(x)$ is
varying slowly along C. This relaxation does not affect the above
conclusion because the local fluctuations of $\phi(x)$ can not distort
the global property happening at $x=+\infty$. For simplicity, we let
$\phi(x)$ jumps $\pi$ at two long separating points and keeps constant
elsewhere. This new layout of $\phi(x)$ is just describing a chain with
a pair of domain-walls, which is chain B actually. So this pair of
domain-walls must take the responsibility of the pair of WFs that have
been lost and gained. Because the two domain-walls are identical through
a mirror reflection, their properties must be the same. So each
domain-wall is in response to {\it one half} of the WF pair. 

{\it The third step.---} With the particular parameters, $\mu=0$,
$\delta=0.2$,$\Delta=0.3$, $B=0.3$ and $\alpha_R=0$, the Hamiltonian
$H(k)$ can be decoupled into two parts, $H_\pm$. The pair of WFs that
have been lost and gained comes from those of $H_-$. So we only need to
focus on the partial Hamiltonian $H_-$, which has been given explicitly.
In this representation, $\alpha_R$ is playing the role of
superconducting pairing. When $\alpha_R=0$,
this Hamiltonian regresses to a standard spinless SSH model. The
dimension of $H_-$ has been extended from that of the standard spinless
SSH model,$2\times 2$, to $4\times4$ because a Nambu representation is
still taken. The two sets of WFs, in blue and in red, come from the
empty conduction band and the filled valence band of the SSH model,
respectively\cite{Xiong.arX}. So in the representation of $H_-$, the
pair of lost and gained WFs corresponds to one quasi-particle
difference.

{\it The fourth step.---} After returning back to the ordinal Nambu
representation of $H(k)$, the one quasi-particle difference between A
and B corresponds to the difference of the parities of fermion numbers
in A and B. Tunning on $\alpha_R$ and $\mu$ does not disturb this
conclusion because the spin-orbital interaction and chemical potential
commute with particle number operator so that they also commute with the
parity. 

Through the above $4$ steps, we have topologically proved that the total
fermion parity on chains A and B, $P_A$ and $P_B$, are different,
$|P_A-P_B|=1$. So each JRBS takes the responsibility of one half of the
parity difference and FP comes out naturally.

We also numerically calculate the parity of the chains A and B with
length $N=400$ and periodic boundary condition. The fermion parity is
calculated by $P= \text{rank}(v) \; {\text{mod}} \, 2$
\cite{arXiv1406.5172}, where $\text{rank}(v)$ is the rank of Bogoliubov
matrix $v$. We confirm that $|P_A - P_B|=1$ in ``TS I'' and $|P_A -
P_B|=0$ elsewhere. 

\subsection{The nonuniversal average charge carried by JRBS} 

We have argued that, when the superconducting pairing is nonzero, JRBS
should not carrying the universal fractional charge $e/2$, because the
particle number is not well defined. We numerically confirm it by
calculating the electric charge $Q$ (in the units of $e$) carried by a
JRBS \cite{PhysRevB.25.6447}, 
\begin{equation}
  Q=\rho_L^{\text{WD}}-\rho_L^0, \label{Qnum} 
\end{equation} 
where $\rho_L^{\text{WD}}$ is the average total particle number in a
segment with a domain-wall at its center and $\rho_L^0$ is the average
particle number for a segment without the domain-wall. $L$ is the length
of these segments which should exceed the localization length of JRBS.
In the numerical calculation, we choose $L=200$ which is long enough for
a saturated $Q$.

\begin{figure}[ht]
  \includegraphics[width=0.45\textwidth]{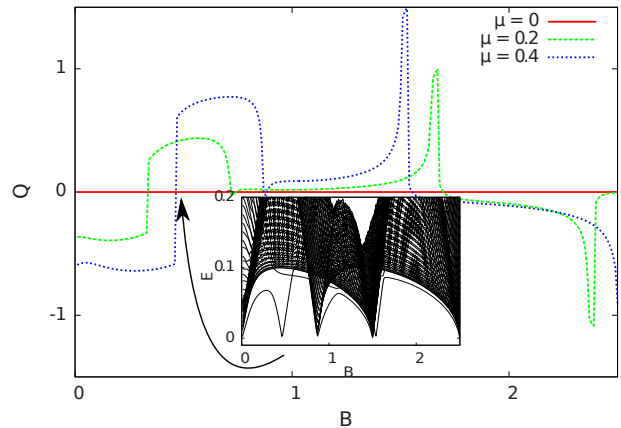}

  \caption{(Color online) Electric charge carried by a domain-wall. The
    system is a ring embedded with a domain-wall with the parameters,
    $\delta=0.2$, $\Delta=0.3$ and $\alpha_R=0.1$. The length of the
    ring is $N=401$ and the length of the segments are $L=200$.
    $\rho_L^{\text{WD}}$ and $\rho_L^0$ are numerically calculated on a
    half of the ring with the domain-wall at the center and on the part
    of the rest half (excluding one site at the end),respectively. The
    inset shows the energy spectrum for the ring with $\mu=0.4$. The
  arrow indicates the consistence of the point at where the zero energy
  crossing happens and electric charge $Q$ switches sign.}

  \label{fig2_7}
\end{figure}

The electric charge $Q$ as a function of $B$ is showed in Fig. \ref{fig2_7}. It
is confirmed that $Q$ becomes non-universal and is
dependent on $\mu$, as well as on $B$ in ``TS I''. When $\mu=0$, the
domain-wall becomes neutral because the particle number on each site is
exactly one, independent of the presence of domain-wall. When $\mu
\ne0$, the nonzero $Q$ is smoothly varying in ``TS I'', except near
a $B_0$ at which its sign is switched. This sign switching is directly
associated with the zero energy crossing for JRBS showed in the inset.
In inset, we show the energy spectrum for the ring with $\mu=0.4$.  The
eigen-energy inside the bulk gap is for the JRBS on domain-wall. It is
the particle-hole transition for the JRBS around the zero energy
crossing point that changes the sign of electrical charge $Q$. 

In ``TS II'', the charge shows a peak and a dip at the phase boundaries.
But it is almost zero in the region.  We suggest that the peak and dip
are due to the quantum fluctuation accompanied with the band gap
closing. 

\subsection{Unavoidable zero energy crossing} 

The energy spectrum in inset of Fig. \ref{fig2_7} (as well as in Fig.
\ref{fig2_2}) shows a zero energy crossing for JRBS. Now we apply a
topological argument to prove that the zero energy crossing is
unavoidable. We start from a proof by contradiction by supposing that
the energy spectrum for JRBS does not cross zero energy. If that is
ture, one can modify factors, i.e., the size of the domain-wall, to
continuously change its eigen-energy from deep-in-gap to near the bulk
band. In this case, the eigenenergy of JRBS is not different from that
of a normal impurity.  When embedding such a domain-wall in a uniform
chain, its contribution of fermion parity is fixed, either $0$ or $1$.
When the embedded domain-walls become two, their total contributions of
fermion parity become $0$. But as we have showed, $|P_A-P_B|=1$, which
requires that the two JRBSs must contribute an extra fermion parity.
Here, we get the contradicting results so that the initial assumption
must be wrong. So the FP JRBS in ``TS I'' must trigger an eigenstate
with its eigenenergy crossing the zero energy inevitably. 

\begin{figure}[ht]
  \includegraphics[width=0.45\textwidth]{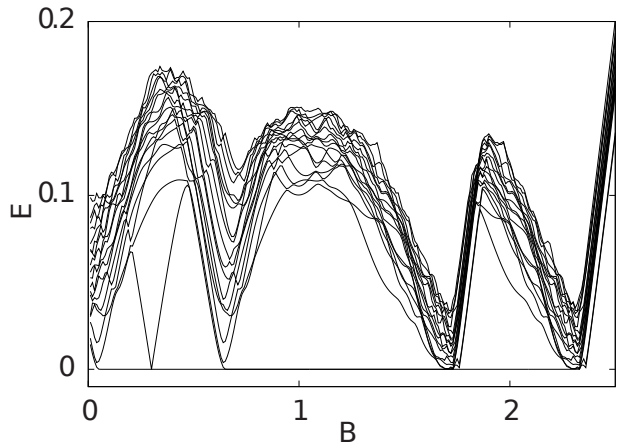}
  \caption{The energy spectrum for a disordered lattice. The parameters
    are the same as those in Fig. \ref{fig2_2}. The disordered system is
    a $N=401$ chain with open boundary condition and embedded by a
    domain-wall at the center.}
  \label{fig2_8}
\end{figure}

One can confirm the robustness of the crossing by studying a disordered
lattice. Here we study a model with the disordered hopping integral
between the nearest neighboring sites, $\delta_i = \delta(1+w_i)$, where
$\delta=0.2$ and $w_i$ is randomly distributing in $[-0.6,0.6]$. The
spectrum is showed in Fig. \ref{fig2_8}. It shows that the zero energy
crossing for the eigenenergy of JRBS is robust against the lattice
distortion. 

\subsection{Majorana Fermion and JRBS} 

In the previous discussion, our focus is on JRBS. In this subsection, we
show the coexistence of MF and JRBS and how to switch between the empty
state and the occupied state of MZES with the help of JRBS. 

\begin{figure}[ht]
  \includegraphics[width=0.45\textwidth]{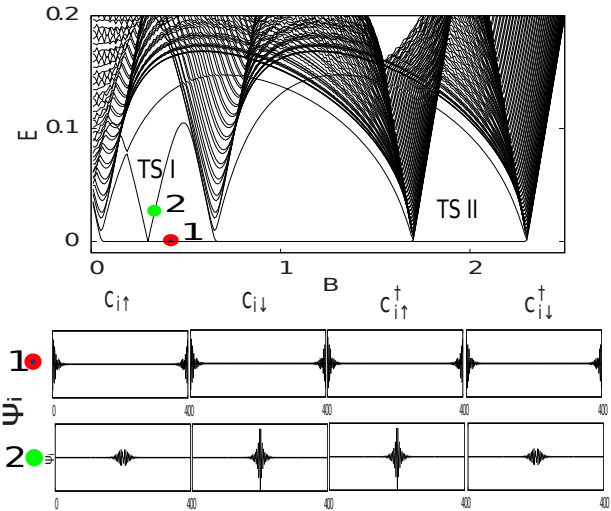}
  \caption{(Color online) Upper panel: energy spectrum for the chain
    embedded with a domain-wall. Lower panel: amplitudes of
    wavefunctions of MZES and JRBS in representation $(c_{i\uparrow},
    c_{i\downarrow}, c^\dagger_{i\uparrow}, c^\dagger_{i\downarrow})^T$,
    where $i$ runs through the lattice sites from $1$ to $N=401$. The
    other parameters are the same as those in Fig. \ref{fig2_7} except
  $\mu=0$. }
  \label{fig2_9}
\end{figure}

In Fig. \ref{fig2_9}, we show the typical energy spectrum for an open
chain embedded with a domain-wall at the center. In ``TS I'', the persistent zero
energy state is MZES and the nonzero eigen-energy of JRBS crosses the
zero energy at $B_0$.  As showed explicitly in the figure, the
wave-functions of these states are localized at the domain-wall for JRBS
and at the geometrical ends for MZES.

When we ignore the MZES by modifying the geometry of the model from
chain to ring (no geometrical ends). It is known that the fermion
parity of ground state of the ring is changed when $B$ is varying across
$B_0$ because of the zero energy crossing. This is confirmed by the
numerical calculation on the parity of the ring. So the ground states on
$B<B_0$ and $B>B_0$ in ``TS I'' have different fermion parity.
Therefore, if we increase $B$ to cross $B_0$ with a ring at its ground
state initially, the final state must be an excited state and can not
spontaneously jump back to the final ground state because the parity is
conserved in this process.

When the MZES is reconsidered in a chain, the above excited state can
jump back to the final ground state by a parity compensation on MZES.
This compensation is achieved by the switching between the empty state
and the occupied state of MZES because this switching contributes one
parity change.  In this manner, with the help of a JRBS embedded in the
chain, we would be able to flip between the two states of the MZES still
in the restriction that the total fermion parity is conserved. A quantum
qubit can be encoded into these two states of one MZES, while in chains
without JRBS, two MZESs are needed. 

\section{The second model with local AF order.}

\begin{figure}[ht]
  \includegraphics[width=0.45\textwidth]{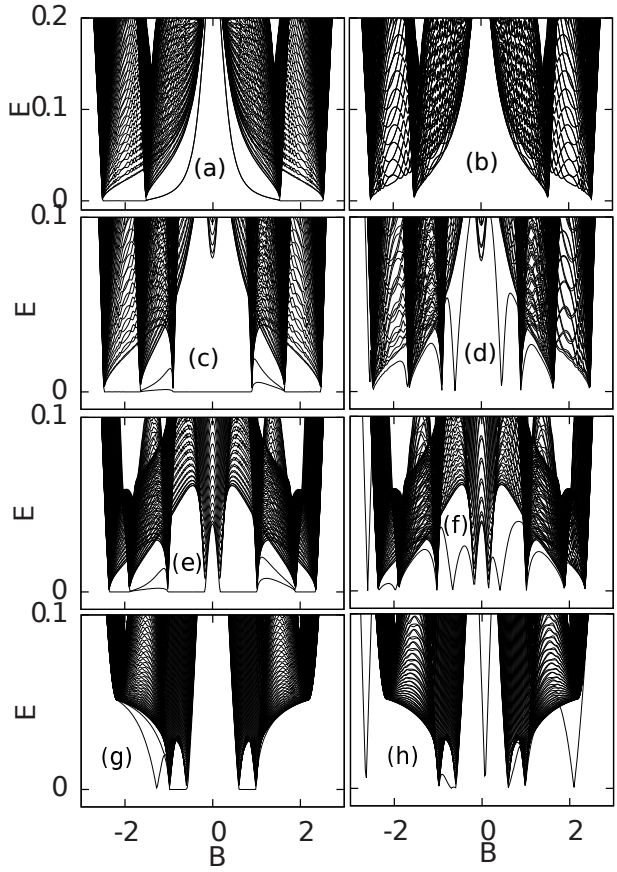}
  \caption{The energy spectrum for a $N=400$ chain with open boundary
    condition (left column) and for a $N=401$ ring with an AF domai-wall
    (right column). The parameters are $\theta=0.6$, $\Delta=0.3$ and
    $\alpha_R=0.1$ in these panels. $M$ is $0$ (a) (b), $0.5$ (c) (d),
    $0.8$ (e) (f) and $1$ (g) (h).} 
  \label{fig3_1}
\end{figure}

The Hamiltonian reads,
\begin{eqnarray}
  H & = & \sum_{i\beta} [(c^\dagger_{i\beta}
  c_{i+1\beta}+\text{h.c.}) +\mu c^\dagger_{i\beta}c_{i\beta}] 
  \nonumber \\  & & +
  i \alpha_R \sum_{i\beta\gamma}
  (c^\dagger_{i\beta}
  \sigma^y_{\beta\gamma} c_{i+1\gamma} +\text{h.c.}) \nonumber \\ & & 
  +\sum_{i\beta\gamma} 
c^\dagger_{i\beta} (\vec{B}-\vec{M}_i)\cdot\vec{\sigma}_{\beta\gamma}
c_{i\gamma}  \nonumber \\ & & +
\Delta \sum_{i}  (c^\dagger_{i\uparrow} c^\dagger_{i\downarrow}
+\text{h.c.}),
  \label{magH}
\end{eqnarray}
where hopping, chemical potential, spin-orbital interaction, Zeeman
interaction caused by a uniform magnetic field $\vec{B}$ and staggered
local magnetic momenta $\vec{M}_i$ and s-wave superconducting pairing
are expressed, respectively. We fix the magnetic field in the $x-z$
plane with $\vec{B}=\sin(\theta) B \hat z +\cos(\theta) B \hat x$ and
the staggered local AF momenta are in the $x$ direction, $\vec{M}_i=
(-1)^i M \hat x$.  

In experiment, AF magnetic order and s-wave superconducting pairing can
be introduced to a 1D semi-conductor through proximity effect by
sandwiching it with AF material and superconductor.

This Hamiltonian in the momentum space in the representation of the
sublattice, the spin and the particle-hole subspaces
,$(\psi_{kA\uparrow}, \psi_{kB\uparrow}, \psi_{kA\downarrow},
\psi_{kB\downarrow}, \psi^\dagger_{-kA\uparrow},
\psi^\dagger_{-kB\uparrow}, \psi^\dagger_{-kA\downarrow},
\psi^\dagger_{-kB\downarrow})^T$, reads
\begin{equation}
  H=\begin{pmatrix} H_0 & V \\ V^\dagger & -H_0 \end{pmatrix},
  \label{Ham0}
\end{equation}
where 
\begin{widetext}
\[
  H_0=\begin{pmatrix} \mu+B\sin(\theta) & 1+e^{-ik} & B\cos(\theta)+M & \alpha_R
    (1-e^{-ik}) \\
    1+e^{ik} & \mu+B\sin(\theta) & -\alpha_R(1-e^{ik}) & B\cos(\theta)-M  \\
    B\cos(\theta)+M & -\alpha_R(1-e^{-ik}) & \mu-B\sin(\theta) & 1+e^{-ik} \\
    \alpha_R(1-e^{ik}) & B\cos(\theta)-M & 1+e^{ik} & \mu-B\sin(\theta)
  \end{pmatrix},
\]
\end{widetext}
and
\[
  V=\begin{pmatrix} 0 & 0 & \Delta & 0 \\ 0 & 0 & 0 & \Delta \\
    -\Delta & 0 & 0 & 0 \\ 0 & -\Delta & 0 & 0 \end{pmatrix}.
\]

When $\mu=0$, the above Hamiltonian can also be decoupled into two
partitioning parts,
\begin{equation}
  H \to \begin{pmatrix} H_+ & 0 \\ 0 & H_- \end{pmatrix},
\end{equation}
where
\begin{widetext}
\[
  H_+=\begin{pmatrix} B\sin(\theta)+\Delta & 1+e^{-ik} & B\cos(\theta)+M & \alpha_R
    (1-e^{-ik}) \\
    1+e^{ik} & B\sin(\theta) -\Delta & -\alpha_R(1-e^{ik}) & B\cos(\theta)-M  \\
    B\cos(\theta)+M & -\alpha_R(1-e^{-ik}) & \Delta-B\sin(\theta) & 1+e^{-ik} \\
    \alpha_R(1-e^{ik}) & B\cos(\theta)-M & 1+e^{ik} &
    -B\sin(\theta)-\Delta
  \end{pmatrix}
\]
\end{widetext}
and $H_-=-H_+$ after a unitary transformation 
\[
  U=\frac{1}{\sqrt{2}}\begin{pmatrix} 
    1 & 0 & 0 & 0 & 0 & 0 & -1 & 0 \\
    0 & 1 & 0 & 0 & 0 & 0 & 0 & 1 \\
    0 & 0 & 1 & 0 & 1 & 0 & 0 & 0 \\
    0 & 0 & 0 & 1 & 0 & -1 & 0 & 0 \\
    0 & 0 & -1 & 0& 1 & 0 & 0 & 0 \\
    0 & 0 & 0 & 1& 0 & 1 & 0 & 0 \\
    1& 0 & 0 & 0 & 0 & 0 & 1 & 0 \\
    0 & -1& 0 & 0 & 0 & 0 & 0 & 1 \\ \end{pmatrix}
\]

So when $\mu=0$, phase transition happens at the points
$M^2=B^2\cos(2\theta)+\Delta^2-4 \pm
2\sqrt{-B^4\cos^2(\theta)\sin^2(\theta) +B^2\Delta^2\cos^2(\theta)
+4B^2\sin^2(\theta)-4\Delta^2}$ and $M^2=B^2\cos(2\theta)
+4\alpha^2_R+\Delta^2-2B\sqrt{-B^2\cos^2(\theta)\sin^2(\theta)
+\Delta^2\cos^2(\theta) -4\alpha^2_R\sin^2(\theta)}$. 

In Fig. \ref{fig3_1} we plot the energy spectrum for a chain with
open boundary condition (left column) and for a ring with one AF
domain-wall on it (right column). Like that in the first model, the AF
domain-wall is simulated by two adjacent $M_i$s pointing to the same
direction.  We find in the cases (c), (d), (e) and (f), there is a TSP
in which a MF zero energy bound state can coexist with a JRBS. These
properties are the same as those showed for the first model.

\begin{figure}[h]
  \includegraphics[width=0.45\textwidth]{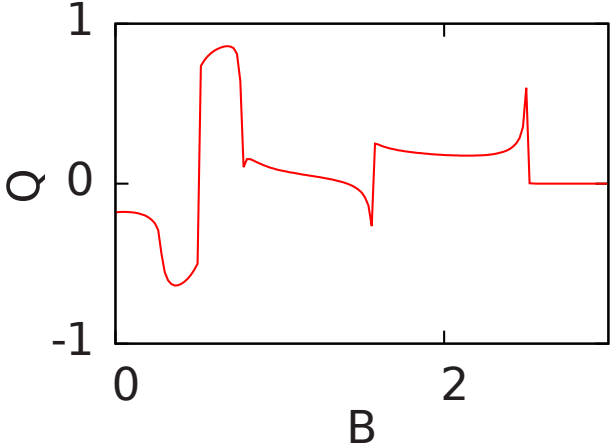}
  \caption{The electric charge carried by an AF domain-wall. The
  parameters are $\theta=0.6$, $\Delta=0.3$, $\alpha_R=0.1$ and $M=0.2$.}
  \label{fig3_2}
\end{figure}

We numerically calculate the electric charge carried by an AF
domain-wall. As Fig. \ref{fig3_2} shows, it is non-universal just as
that in the first model. We also numerically calculate the parities of
fermion numbers for chains like A and B. The result is same as that in
the first model. So the JRBS attached to the AF domain-wall in this
model is also carrying FP. As we have discussed in the previous model,
this means that the eigenenergy of JRBS must suffer an unavoidable zero
energy crossing.


\section{Conclusions}

We have showed that JRBS and MF can coexist in a
TSP in 1D models. The eigen-energy of the FP JRBS suffers an unavoidable
zero energy crossing.  This crossing separates the TSP into two parts
with different parities for the ground state. This can be used to switch
between the occupied and the empty states of MZES under the conservation
of total fermion parity. One should be able to observe such effect by
measuring the Josephson current through MZES. As the magnetic
field is modified across the crossing point, the Josephson current
should suffer a sudden sign jump because the parity on the MZES is
changed.  It still remains challenging how to experimentally observe the
FP JRBS directly. One possible way is to apply the proposal in Ref.
\cite{PhysRevLett.112.196803}, although the electric charge on the
domain-wall is not $e/2$ in this case.

{\it Acknowledgments.---} 
The work was supported by the State Key Program for Basic Research of
China (Grant Nos. 2009CB929504, 2009CB929501), National Foundation of
Natural Science in China Grant Nos. 10704040, 11175087.


\end{document}